\documentclass[fleqn,twoside]{article}
\usepackage{espcrc2}
\usepackage{graphicx}

\usepackage{amssymb}

\newcommand{\fun}{$F_2^{\gamma}(x,Q^2)$ }

\newcommand{\gam}{^{\gamma}}
\newcommand{\fund}{$F_2^{\gamma}(x,Q^2)$}
\newcommand{\be}{\begin{equation}}
\newcommand{\ba}{\begin{eqnarray}}
\newcommand{\ea}{\end{eqnarray}}
\newcommand{\etal}{{\it et al.}}

\newcommand{\AmS}{{\protect\the\textfont2
  A\kern-.1667em\lower.5ex\hbox{M}\kern-.125emS}}

\def\eg{{e.g.} }
\def\ie{{i.e.} }

\title{
CJK- Improved LO Parton Distributions in 
the Real Photon and Their Experimental Uncertainties}

\author{F. Cornet \address[GU]{Departamento de F\'{\i}sica Te\'orica y del 
        Cosmos, Universidad de Granada, \\ 
        Campus de Fuente Nueva, E-18071, Granada, Spain},
        P. Jankowski\address[UW]{Institute of Theoretical Physics, Warsaw 
        University, \\
        ul. Ho\.za 69, 00-681 Warsaw, Poland}
        and
        M. Krawczyk\addressmark[UW]}

\begin{document}


\thispagestyle{empty}

\begin{flushright}
{ \large ITF-2003-28 \\
UG-FT-155/03 \\ 
CAFPE-25/03 \\ }
\end{flushright}

\vskip 1.5cm

\begin{center} 
\Large{ CJK- Improved LO Parton Distributions in 
the Real Photon and Their Experimental Uncertainties}

\vskip 1cm

\large{F. Cornet}

\vskip 0.4cm

{\sl Departamento de F\'{\i}sica Te\'orica y del Cosmos, Universidad de 
     Granada, \\ 
     Campus de Fuente Nueva, E-18071, Granada, Spain}

\vskip 1cm

\large{P. Jankowski and M. Krawczyk}

\vskip 0.4cm

{\sl Institute of Theoretical Physics, Warsaw University, \\
     ul. Ho\.za 69, 00-681 Warsaw, Poland}

\vskip 1cm

\end{center}

A new analysis of the radiatively generated, LO quark ($u,d,s,c,b$) and gluon 
densities in the real, unpolarized photon, improved in respect to our paper 
\cite{cjkl}, is presented. We perform four new global fits to the experimental
data for $F_2^{\gamma}$, two using a standard FFNS approach and two based on 
ACOT$_{\chi}$ scheme \cite{acot}, leading to the FFNS$_{CJK}$ and CJK models. 
We also present the analysis of the uncertainties of the new CJK 2 parton 
distributions due to the experimental errors, the very first such analysis 
performed for the photon. This analysis is based on the Hessian method, for a 
comparison for chosen cross-sections we use also the Lagrange method.

\clearpage


\pagenumbering{arabic}

\begin{abstract}
A new analysis of the radiatively generated, LO quark ($u,d,s,c,b$) and gluon 
densities in the real, unpolarized photon, improved in respect to our paper 
\cite{cjkl}, is presented. We perform four new global fits to the experimental
data for $F_2^{\gamma}$, two using a standard FFNS approach and two based on 
ACOT$_{\chi}$ scheme \cite{acot}, leading to the FFNS$_{CJK}$ and CJK models. 
We also present the analysis of the uncertainties of the new CJK 2 parton 
distributions due to the experimental errors, the very first such analysis 
performed for the photon. This analysis is based on the Hessian method, for a 
comparison for chosen cross-sections we use also the Lagrange method. 
\vspace{1pc}
\end{abstract}

\maketitle


\section{Introduction}

We continue our LO analysis \cite{cjkl} of the parton distributions in the 
unpolarized real photon, which led us to a construction of the CJKL parton 
parametrization, improving and broadening our investigation. Like previously 
our aim is to develop a proper description of the heavy charm- and 
bottom-quark contributions to the photon structure function, \fund. The new 
models FFNS$_{CJK}$ 1 and CJK 1 are slightly modified versions of our previous
FFNS$_{CJKL}$ and CJKL models, respectively. In addition we analyse the 
FFNS$_{CJK}$ 2 model, which includes the so-called ``resolved-photon'' 
contribution of heavy quarks to \fun, given by the $\gamma^* G \to h\bar h$ 
process, and CJK 2 model, with an improved positivity constraint. All models 
are based on the idea of radiatively generated parton distributions introduced
by the GRV group (see \eg \cite{grv92}).

In this analysis we neglect TPC2$\gamma$ data, as in other recent analyses,
and slightly modify, with respect to previous analysis, both types of models. 
Moreover, for the very first time for the photon, we estimate uncertainties of
the parton distributions due to the experimental errors of data. Following the
analyses of this type performed for proton structure by the CTEQ Collaboration,
\cite{cteq1}--\cite{cteq3} and the MRST group, \cite{mrst} we use the Hessian 
method to obtain sets of parton parametrizations allowing, along with the 
parton parametrization of the best fit, to calculate the best estimate and 
uncertainty of any observable depending on the parton densities in real 
photon. We compare our results for $F_{2,c}^{\gamma}$ and prompt photon 
production in $\gamma \gamma$ obtained with the Hessian and Lagrange
(\cite{cteq1},\cite{mrst}, \cite{cteq4}) methods.


\section{FFNS$_{CJK}$ and CJK models}

The difference between the FFNS$_{CJK}$ and CJK models lays in the approach to
the calculation of the heavy quark $h$ ($c$ and $b$) contributions to the 
photon structure function \fund. The FFNS$_{CJK}$ models base on a widely 
adopted Fixed Flavour Number Scheme in which there are no heavy quarks among 
partons. They contribute to \fun by a 'direct' (Bethe-Heitler) 
$\gamma^* \gamma \to h\bar h$ process ($F_{2,h}\gam(x,Q^2)|_{dir}$), in 
addition one can also include the so-called 'resolved'-photon contribution
from  $\gamma^* G \to h\bar h$ ($F_{2,h}\gam(x,Q^2)|_{res}$) \cite{GRSt}:
\ba
F_2\gam(x,Q^2)|_{FFNS} =x \sum_{i=1}^3 e_i^2 (q_i\gam +\bar q_i\gam)(x,Q^2)\\
+ \sum_{h(=c,b)}^2 \left[ F_{2,h}^{\gamma}(x,Q^2)|_{dir} + 
F_{2,h}\gam(x,Q^2)|_{res} \right], \nonumber
\ea
with $q\gam_i$ ($\bar q\gam_i$) being the light $u,d,s$ quark (anti-quark)
densities governed by the DGLAP evolution equations.

The CJK models adopt the ACOT$_\chi$ scheme \cite{acot} which is a recent 
realization of the Variable Flavour Number Scheme. In this approach heavy 
quarks are, similarly to the light quarks, constituents of the photon. 
Therefore, apart from the direct and resolved-photon contributions  also 
$q_h$ contribute to \fun. It leads to a double counting of the heavy quark 
contributions to \fun, therefore inclusion of the corresponding subtraction 
terms, $F_{2,h}^{\gamma}|_{\mathrm{dir},\mathrm{sub}}$ and 
$F_{2,h}^{\gamma}|_{\mathrm{res},\mathrm{sub}}$, is necessary.

Next, following the ACOT$_\chi$ approach, we use the $\chi_h$ parameters to 
obtain the proper vanishing of the heavy-quark densities below the kinematic 
thresholds, for DIS given by the $W^2=Q^2(1-x)/x=4m_h^2$. Substitution of $x$ 
with $\chi_h=x(1+4m_h^2/Q^2)$ in $q_h$ and in subtraction terms should lead 
to the correct threshold behaviour as $\chi_h \to 1$ for $W \to 2m_h$. 
In the CJK models 
\ba
F_2^{\gamma}(x,Q^2)|_{CJK} = \sum_{i=1}^3 xe_i^2 (q_i\gam +\bar q_i\gam)(x,Q^2) \\
+ \sum_{h(=c,b)}^2 xe_h^2 (q_h\gam + \bar q_h\gam)(\chi_h,Q^2) \nonumber \\
+ \sum_{h(=c,b)}^2 \left[ F_{2,h}^{\gamma}(x,Q^2)|_{dir} + 
F_{2,h}^{\gamma}(x,Q^2)|_{res} \right] \nonumber \\
- \sum_{h(=c,b)}^2 \left[ F_{2,h}^{\gamma}|_{\mathrm{dir},\mathrm{sub}}
+ F_{2,h}^{\gamma}|_{\mathrm{res},\mathrm{sub}} \right]. \nonumber
\ea

The QCD evolution (DGLAP) starts from a scale chosen to be small, $Q_0^2=0.25$
GeV$^2$, for both types of models, hence our parton densities are radiatively 
generated. As it is well known the point-like solutions of the evolution 
equations are calculable without further assumptions, while the hadronic parts 
need the input distributions. For this purpose we utilize the Vector Meson 
Dominance (VMD) model, assuming
\be
f_{had}\gam(x,Q_0^2) = \sum_{V}\frac{4\pi \alpha}{\hat f^2_{V}}f^{V}(x,Q_0^2)=
\kappa\frac{4\pi \alpha}{\hat f^2_{\rho}}f^{\rho}(x,Q_0^2), 
\end{equation}
where the sum over all light vector mesons (V) is proportional to the 
$\rho^0$-meson parton density with a parameter $\kappa$. We take the input 
densities of the $\rho$ meson at $Q_0^2$ in {the} form of valence-like 
distributions both for the (light) quarks ($v^{\rho}$) and gluons ($G^{\rho}$):
\ba
xv^{\rho}(x,Q_0^2) &=& N_v x^{\alpha}(1-x)^{\beta},  \\
xG^{\rho}(x,Q_0^2) &=& \tilde N_g xv^{\rho}(x,Q_0^2)= 
N_g x^{\alpha}(1-x)^{\beta}, \nonumber \\
x \zeta^{\rho}(x,Q_0^2)&=&0, \nonumber
\label{input1}
\ea
where $N_g=\tilde N_gN_v$. All sea-quark distributions (denoted as 
$\zeta^{\rho}$), including $s$-quarks, are neglected at the input scale. 
The valence-quark and gluon densities satisfy the constraint 
representing the energy-momentum sum rule for $\rho$:
\be
\int_0^1 x(2v^{\rho}(x,Q_0^2)+G^{\rho}(x,Q_0^2))dx = 1.
\label{const2}
\end{equation}
One can impose an extra constraint related to the number of valence quarks:
\begin{equation}
n_v = \int_0^1 2 v^\rho (x,Q_0^2) = 2.
\label{nv}
\end{equation}
Use of both constraints allows to express $N_v$ and $N_g$ in terms of 
$\alpha$, $\beta$ and $\kappa$, reducing the number of free parameters to 
three.


\section{New analysis}

In our new analysis we improve treatment of the running of $\alpha_s$
by differentiating the number of active quarks in the running of $\alpha_s$ 
and in the evolution equations, and by using lower values of $\Lambda_{QCD}$.
We first describe new aspects of our analysis which are common to 
all considered models.


\subsection{Data}

New fits were performed using all, \fun data, except the old TPC$\gamma$.
In our former global analysis \cite{cjkl} we used 208 \fun experimental 
points. Now we decided to exclude the TPC2$\gamma$ data (as for instance 
in \cite{klasen}) since these data are considered to be not in agreement with 
other measurements. After the exclusion we have 182 data points.


{\subsection{$\alpha_s$ running and values of $\Lambda^{(N_q)}$}}

The running of the strong coupling constant at lowest order is given 
by the well-known formula:
\be
\alpha_s^{(N_q)}(Q^2) = \frac{4\pi}{\beta_0 \ln (Q^2/\Lambda^{{(N_q)}^2})}, 
\; \beta_0 = 11-\frac{2}{3}N_q,
\label{alphasold}
\end{equation}
where $\Lambda^{(N_q)}$ is the $\Lambda_{QCD}$ value for $N_q$ 
quarks. $N_q$ increases by one whenever $Q^2$ reaches a heavy quark threshold,
\ie when  $Q^2 = m_h^2$. The condition 
$\alpha_s^{(N_q)}(m_h^2)=\alpha_s^{(N_q+1)}(m_h^2)$ is imposed in order to 
ensure the continuity of the strong coupling constant. In the new analyses we 
introduce the number of active quarks in the photon, denoted by $N_f$, which 
differs from the number of quarks contributing to the running of $\alpha_s$.

The distinction between both numbers of quarks forces to use slightly more 
complicated formulae for the evolution of the parton densities than in our 
previous analysis. More precisely, we must proceed in three steps to perform 
the DGLAP evolution. In the first step, describing the evolution from the 
input scale $Q_0$ to the charm-quark mass $m_c$, the hadronic input 
$q_{\mathrm{had}}\gam(x,Q_0^2)$ is taken from the VMD model. In the second 
step we evolve the parton distributions from $m_c$ to the beauty-quark mass, 
$m_b$, a new hadronic input is given by the sum of already evolved hadronic 
and point-like contributions to the parton density. The point-like 
distribution at $Q^2=m_c^2$ becomes zero again. The same is repeated for 
$Q^2>m_b^2$. That way we can solve the equations for three ranges of $Q^2$, in
which $N_f=3,4$ and 5, separately. In each range values of $b_0$ and 
$\Lambda$ depend on $N_q$.

In the previous work we {assumed} ({following the GRV group approach 
\cite{grv92}}) that the LO and NLO $\Lambda$ values for four active 
flavors are equal, and used $\Lambda^{(4)}=280$ MeV from the Particle Data 
Group {(PDG)} report \cite{prd}. We now abandon this assumption and take for 
$\Lambda^{(4)} = 115$ MeV, which is obtained in LO from the world average 
value $\alpha_s(M_Z) = 0.117$, with $M_Z = 91.188$ GeV, using 
Eq. \ref{alphasold}. Imposing the continuity condition for the strong coupling
constant and $m_c = 1.3$ GeV and $m_b = 4.3$ GeV, we obtain 
$\Lambda^{(3)} = 138$ MeV and $\Lambda^{(5)} = 84$ MeV.


\subsection{VDM}

In our new analysis we try to relax the constraint on $n_v$ (\ref{nv}). This 
leads to 4-parameter fit.


\subsection{FFNS}

In the FFNS$_{CJK}$ 2 model we include the so-called ``resolved-photon'' 
contribution of heavy quarks to \fun, given by the $\gamma^* G \to h\bar h$ 
process.


\subsection{Subtraction terms in CJK models}

In \cite{cjkl} we derived the  subtraction term for a direct contribution,
$F_{2,h}^{\gamma}|_{dir,subtr}$, from the integration up to $Q^2$ of a part 
of the DGLAP evolution equations, namely: 
\be
\frac{dq_h\gam (x,Q^2)}{d\ln Q^2} = \frac{\alpha}{2\pi}e_h^2 k(x),
\end{equation}
where $k(x)$ is the lowest order photon-quark splitting function (see Eq. (7)
in Ref. \cite{cjkl}). For the lower limit we took in Ref. \cite{cjkl} the 
natural for a Bethe-Heitler process limit $Q^2_{low} = m_h^2$. However, since 
the threshold condition is $W^2 \geq 4 m_h^2$, even for $Q^2 < m_h^2$ the 
heavy-quark contributions do not vanish as long as the condition
$x < Q^2 / (Q^2+ 4 m_h^2)$ is fulfilled. In this paper we take
$Q^2_{low}= Q^2_0$, which improves quality of the fits. The direct subtraction 
term has now the form:
\be
F_{2,h}^{\gamma}|_{dir,subtr}(x,Q^2)= x \ln \frac{Q^2}{Q_0^2}  
3e_h^4 \frac{\alpha}{\pi}(x^2+(1-x)^2).
\label{dirsub2}
\end{equation}

We apply the same change to the subtraction term for the resolved-photon  
contribution:
\begin{equation}
F_{2,h}^{\gamma}|_{res,subtr}= x \ln \frac{Q^2}{{Q^2_0}}
e_h^2 \frac{\alpha_s(Q^2)}{\pi}\int_{x}^1 \frac{dy}{y}
P_{qG}(\frac{x}{y})G\gam(y,Q^2).
\label{ss19}
\end{equation}

As we noticed in Section 2 the $x\to \chi_h$ substitution leads to the 
proper threshold behavior of all the heavy-quark contributions to the \fund,
except for both subtraction terms. This is already seen in Eq. (\ref{dirsub2})
that this term does not vanish for $\chi_h \to 1$ and therefore by subtracting
it the resulting heavy-quark contribution to $F_2^{\gamma}$ may become 
negative for large $x$. An extra constraint to avoid this unphysical situation 
is, thus, needed. In Ref. \cite{cjkl} we imposed the simple condition 
(positivity constraint) that the heavy quark contribution to \fun has to be 
positive. Unfortunately, this constraint was not strong enough and for some 
small windows at small and large $x$ the still unphysical situation 
$F_{2,h}\gam(x,Q^2)<F_{2,h}\gam(x,Q^2)|_{direct}+
F_{2,h}\gam(x,Q^2)|_{resolved}$ 
was found \cite{MariuszPrzybycien}. In this paper we apply new condition:
\be
F_{2,h}\gam(x,Q^2) \geq F_{2,h}\gam(x,Q^2)|_{dir} + F_{2,h}\gam(x,Q^2)|_{res}.
\label{condnew}
\end{equation}


\section{Results of the new $F_2^{\gamma}$ global fits}

In the fits we use 182 \fun experimental points with equal weights. Fits 
based on the least-squares principle (minimum of $\chi^2$) were done using 
\textsc{Minuit} \cite{minuit}. Systematic and statistical errors on data
points were added in quadrature.

The CJK 1 and 2 models differ in the form of the positivity constraint.
In the CJK 1 model we keep the old condition $F_{2,h}\gam(x,Q^2)>0$ while the 
CJK 2 model imposes the condition (\ref{condnew}). In these models we do not 
apply the sum rule (\ref{nv}) and have four free parameters: 
$\alpha,\beta,N_v,\kappa$ related to the initial quark and gluon densities at 
the scale $Q^2_0=0.25$ GeV$^2$ (\ref{input1}).

The two FFNS models differ by the resolved-photon contribution of heavy quarks 
to \fund. It appears only in the FFNS$_{CJK}$ 2 model. In both FFNS models we 
impose constraints (\ref{const2}) and (\ref{nv}). That allows to write $N_v$ 
in terms of $\alpha$, $\beta$ and $\kappa$ reducing the number of free 
parameters to three \footnote{The test fits without the number of valence 
quarks constraint gave $n_v \approx 0.5$ and $\approx 1.4$ in the 
FFNS$_{CJK}$ 1 and FFNS$_{CJK}$ 2 models, respectively. Both these values
are very far away from the expected value 2.}.

The results for the total $\chi^2$ for 182 points and the $\chi^2$ per degree 
of freedom for our new fits are presented in table \ref{tparam}. The fitted 
values for parameters $\alpha$, $\beta$, $\kappa$ and $N_v$ are also presented 
together with the errors obtained from \textsc{Minos} program with the 
standard requirement of $\Delta \chi^2 = 1$. In the case of the FFNS$_{CJK}$ 
models the $N_v$ parameter is calculated from the constraint (\ref{const2}) 
and therefore we do not state its error. Note, that the valence number 
integral $n_v$ (\ref{nv}) gives in CJK models 1.94 and 2.00, for CJK 1 and 
CJK 2, respectively.

\begin{table*}[htb]
\begin{center}
\renewcommand{\arraystretch}{1.2}
\begin{tabular}{|c|@{} p{0.1cm} @{}|c|c|@{} p{0.1cm} @{}|c|c|c|c|@{} p{0.1cm} @{}|c|}
\hline
 model           && $\chi^2$ (182 pts) & $\chi^2/_{\rm DOF}$ && $\kappa$ & $\alpha$ & $\beta$ & $N_v$ && $ \tilde N_{g}$ \\
\hline
\hline
FFNS$_{CJK}$ 1 &&        314.0       &         1.754       &&  $2.267^{+0.063}_{-0.072}$ & $0.265^{+0.038}_{-0.032}$ & $0.792^{+0.189}_{-0.149}$ & 0.358 && 5.02  \\
\hline
FFNS$_{CJK}$ 2 &&        279.8       &         1.563       &&  $2.110^{+0.084}_{-0.090}$ & $0.310^{+0.054}_{-0.051}$ & $0.823^{+0.265}_{-0.223}$ & 0.415 && 4.51  \\
\hline
 CJK 1         &&        273.9       &         1.539       &&  $2.146^{+0.154}_{-0.144}$  & $0.218^{+0.070}_{-0.063}$ & $0.462^{+0.261}_{-0.227}$ & $0.269^{+0.076}_{-0.059}$ && 5.00  \\
\hline
 CJK 2         &&        273.7       &         1.537       &&  $1.934^{+0.131}_{-0.124}$  & $0.299^{+0.077}_{-0.069}$ & $0.898^{+0.316}_{-0.275}$ & $0.404^{+0.116}_{-0.088}$ && 4.93  \\
\hline
\end{tabular}
\caption{The total $\chi^2$ for 182 data points and per the degree of freedom,
and parameters of the best fits for FFNS$_{CJK}$ and CJK models. Errors are 
obtained from \textsc{Minos} with  $\Delta \chi^2 = 1$.}
\label{tparam}
\end{center}
\end{table*}

The $\chi^2$ per degree of freedom obtained in our new fits, between 1.5 and 
1.7, is better than the old results, mostly due to much lower
$\Lambda^{(N_q)}$ as well as due to modification of the subtraction 
contributions in the CJK models. The old $\chi^2/_{\rm DOF}$ 
for the same set of 182 data points read 1.99 in the $FFNS_{CJKL}$ and 1.80 
in the CJKL model \footnote{Note, however that in old analysis global fits were
performed for 208 points.}.

We observe that the only real difference in $\chi^2/_{\rm DOF}$ is between the
first and other three fits and is a result of inclusion or not of the
$\gamma^* G \to h\bar h$ contribution to \fund. It is obvious that taking this
process into account improves the agreement between model and data.

In light of these results one can conclude that in both, so different  
treatment of heavy-quark contributions to the photon structure function, all 
fitted parameters are  similar. This is related to the fact, that we use in 
global fits only data for $F_2^{\gamma}$, quantity dominated by the 
light-quark contributions.


\subsection{Comparison with the $F_2\gam$ data}

The FFNS$_{CJK}$ 2 and both CJK models predict a much steeper behaviour of the
\fun at small $x$ with respect to FFNS$_{CJK}$ 1 fit and GRS LO \cite{grs} and
SaS1D \cite{sas} parametrizations. On the other hand these curves are less 
steep than the FFNS$_{CJKL}$ and CJKL results from \cite{cjkl}. In the region 
of $x \gtrsim 0.1$, the behaviour of the \fun obtained from all fits and 
parametrizations apart from the CJK 1 model is similar. The shape of the CJK 1 
fit at high $x$ is a result of the $F_{2,h}^{\gamma}(x,Q^2)>0$ condition which
allows for $F_{2,h}^{\gamma}(x,Q^2)<F_{2,h}^{\gamma}(x,Q^2)|_{direct}+F_{2,h}^{\gamma}(x,Q^2)|_{resolved}$. It effects in the lower position of the 
characteristic point at which the charm-quark contributions to the \fun
appear as compared to other models predictions. Apart from that the CJK 1 fit 
gives smaller \fun values around this charm-quark threshold. Finally at high 
$Q^2$ and high $x$ the CJK 1 model produces much lower structure function 
values than all other fits and parametrizations predict. We found that the 
FFNS$_{CJK}$ 1 fit gives very similar prediction to the GRS LO 
parametrization results in the whole range of $x$.

\begin{figure}[h]
\includegraphics[scale=1.0]{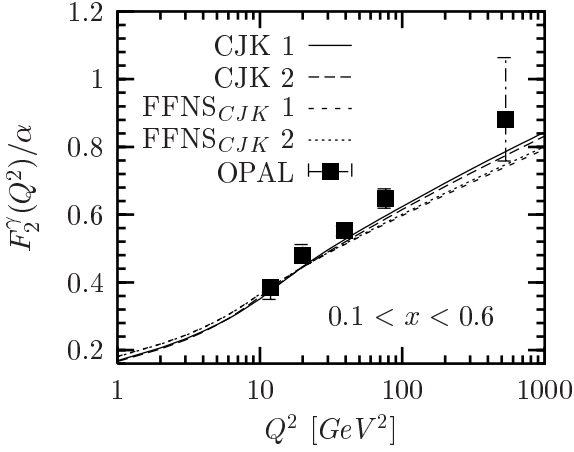}
\vskip -1.3cm
\caption{Comparison of the recent OPAL data \cite{HQ2} for the
$Q^2$-dependence of the averaged over $0.1<x<0.6$ $F_2^{\gamma}/\alpha $
with the predictions of the CJK and FFNS$_{CJK}$ models.}
\label{evol1}
\end{figure}

Figure \ref{evol1} presents the predictions for \fund, averaged over medium 
$x$ regions, compared with the recent OPAL data \cite{HQ2}, not used directly 
in our analysis. Like in our previous analysis we see that all FFNS type 
predictions (including GRS LO and SaS1D parametrizations) are similar and 
fairly well describe the experimental data. The CJK models alike the CJKL 
model gives slightly better agreement with the data.


\subsection{Parton densities}

The parton densities obtained in the CJK and FFNS$_{CJK}$ models are all very 
similar. Of course there are no heavy-quark distributions in FFNS models. In 
case of the CJK models the $c^{\gamma}(x,Q^2)$ and $b^{\gamma}(x,Q^2)$ 
densities vanish not at $x=1$, as for the GRV LO and SaS1D parametrizations, 
but at the kinematical threshold, which was our aim. We notice that our new 
parton densities have all similar shapes but slightly higher values than the 
corresponding old CJKL distributions. In case of the gluon density we find 
that all new curves are much steeper than the predictions of the old models 
and the GRV LO and SaS1D parametrizations, see also below.


\subsection{Comparison with $F\gam_{2,c}$}

\begin{figure}[h]
\includegraphics[scale=1.0]{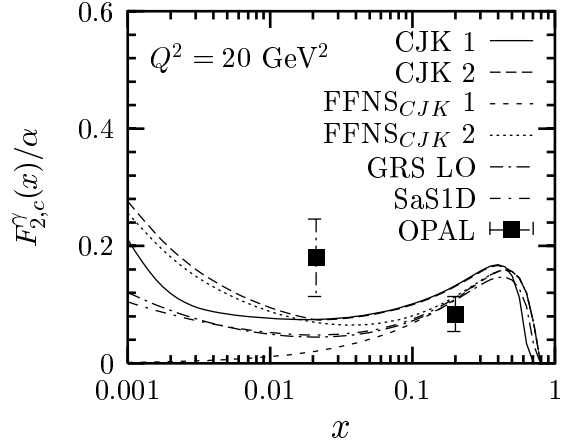}
\vskip -1.7cm
\caption{Comparison of the structure function $F_{2,c}^{\gamma}(x,Q^2)/\alpha$
calculated in the CJK and FFNS$_{CJK}$ models and in GRS LO \cite{grs} and 
SaS1D \cite{sas} parametrizations with the OPAL data \cite{F2c}.} 
\label{fF2c}
\end{figure}

In Fig. \ref{fF2c} we present our predictions for the $F_{2,c}\gam$, 
in comparison with OPAL data \cite{F2c}, not used directly in the analysis,
and results of the GRS LO and SaS1D parametrizations.

All our models containing the resolved-photon contribution (FFNS$_{CJK}$ 2 and
both CJK models) agree better with the low $x$ experimental point than other 
predictions. The GRS LO and SaS1D parametrizations also include the 
resolved-photon term but in their case the gluon density increased less steep 
than our models predict, as was already mentioned. Their $F_{2,c}\gam$ lines 
lie below results of our new fits but higher than the FFNS$_{CJK}$ 1 curve 
which is given solely by the direct Bethe-Heitler contribution.

The CJK models overshoot the experimental point at high $x$ while other 
predictions agree with it within its uncertainty bounds. Again the 
$F_{2,c}\gam$ from the CJK 1 fit vanishes at lower $x$ than in other models.


\section{Uncertainties of the parton distributions}

Following the corresponding analyses for the proton we consider the 
experimental uncertainties of the CJK parton densities. This is first 
analysis of this type for the photon, details are described in \cite{newart}.


\subsection{The Hessian method}

The CTEQ Collaboration \cite{cteq1}-\cite{cteq3}, developed and applied an 
improved method of the treatment of the experimental data errors. Later the 
same formalism has been applied by the MRST group in \cite{mrst}. The method 
bases on the Hessian formalism and as a result one obtains a set of
parametrizations allowing for the calculation of an uncertainty of any 
physical observable depending on the parton densities.

We apply the method as described in Refs. \cite{cteq1} and \cite{cteq2}. 
For the sake of clearness of our procedures we will partly repeat it here 
keeping the notation introduced by the CTEQ Collaboration.

Let us consider a global fit to the experimental data based on the 
least-squares principle performed in a model, being parametrized with a set of 
$\{a_i, i=1,2, \cdots d \}$ parameters. Each set of values of these parameters
constitutes a test parametrization $S$. The set of the best values of 
parameters $\{a_i^0\}$, corresponding to the minimal $\chi^2$, $\chi_0^2$, is 
denoted as $S^0$ parametrization. In the Hessian method one makes a basic 
assumption that the deviation of the global fit from $\chi_0^2$ can be 
approximated in its proximity by a quadratic expansion in the basis of 
parameters $\{a_i\}$
\be
\Delta \chi^2 = \chi^2-\chi_0^2 = \sum_{i=1}^d \sum_{j=1}^d H_{ij}(a_i-a_i^0)
(a_j-a_j^0),
\end{equation}
where $H_{ij}$ is an element of the Hessian matrix, calculated as
\be
H_{ij} = \frac{1}{2}\left( \frac{\partial^2\chi^2}{\partial a_i \partial a_j} 
\right)_{a_0}.
\label{hess}
\end{equation}

Since the $H_{ij}$ is a symmetric matrix, it has a complete set of 
$k=1,2...d$ orthonormal eigenvectors $(v_i)_k$ defined by
\ba
\sum_{j=1}^d H_{ij}(v_{j})_k &=& \epsilon_k (v_{i})_k, \\
\sum_{i=1}^d (v_{i})_j (v_{i})_k &=& \delta_{jk},
\ea
with $\{\epsilon_k\}$ being the corresponding eigenvalues. Variations around
the minimum can be expressed in terms of the basis provided by the set of
eigenvectors
\be
a_i-a_i^0 = \sum_{k=1}^d  s_k z_k (v_{i})_k,
\label{twobasis}
\end{equation}
where $\{z_k\}$ are new parameters describing the displacement from the best 
fit. The $\{s_k\}$ are scale factors introduced to normalize $\{z_k\}$ in 
such a way that
\be
\Delta \chi^2 = \chi^2-\chi_0^2 = \sum_{k=1}^d z_k^2.
\label{quadrz}
\end{equation}
The above equation means that the surfaces of constant $\chi^2$ are spheres in
the  $\{z_k\}$ space. That way the $\{z_k\}$ coordinates create a very useful,
normalized basis. The $(v_i)_k\equiv v_{ik}$ matrix describes the 
transformation between this new basis $\{z_k\}$ and old $\{a_i\}$ basis.
The scaling factors $s_k$ are equal to $\sqrt{1/\epsilon_k}$  in the 
ideal quadratic approximation (\ref{hess}).

The Hessian matrix can be calculated from its definition in Eq. (\ref{hess}). 
Such computation meets many practical problems arising from the large range 
spanned by the eigenvalues $\{\epsilon_k\}$, the numerical noise and 
non-quadratic contributions to $\chi^2$. The solution (an iteration procedure)
has been given by the CTEQ Collaboration \cite{cteq1}.

Having calculated the eigenvectors, eigenvalues and scaling factors we can
create a basis of the parametrizations of the parton densities, 
$\{S_k^{\pm}$, $k=1,\cdots,d\}$. Each parametrization corresponds to a $k$-set 
of ${a_i}$ parameters defined by displacements from $a_i^0$ of a magnitude $t$
``up'' or ``down'' along the corresponding eigenvector direction
\be
a_i(S_k^{\pm}) = a_i^0 \pm t \: (v_{i})_k s_k.
\label{delpar}
\end{equation}
For each $S_k^{\pm}$ parametrization $\Delta \chi^2 = t^2$.

The best value of a physical observable $X$ depending on the photon parton 
distributions is given as $X(S^0)$. The uncertainty of $X$, for a displacement
from the parton densities minimum by $\Delta \chi^2 = T^2$ ($T$ - the 
tolerance parameter), can be calculated from a very simple expression 
(a master equation)
\be
\Delta X = \frac{T}{2t}\left( \sum_{k=1}^d[X(S_k^+)-X(S_k^-)]^2 \right)
^{\frac{1}{2}}.
\label{master}
\end{equation}
Note that having calculated $\Delta X$ for one value of the tolerance parameter
$T$ we can obtain the uncertainty of $X$ for any other $T$ by simple scaling 
of $\Delta X$. This way sets of $\{S_k^{\pm}\}$ parton densities give us a 
perfect tool for studying of the uncertainties of other physical quantities. 
One of such quantities can be the parton densities themselves.

Finally, we can calculate the uncertainties of the $a_i$ parameters of the 
model. According to Eq. (\ref{delpar}) in this case 
$a_i(S_k^+)-a_i(S_k^-) = 2t (v_{i})_k s_k$ and the master equation gives 
a simple expression
\be
\Delta a_i = T\left( \sum_{k=1}^d v_{ik} s_k \right)^{\frac{1}{2}}.
\label{maspar}
\end{equation}

In practice we observe the considerable deviations from the ideal 
quadratic approximation of equation (\ref{quadrz}). To make an improvement we 
can adjust the scaling factors $\{s_i\}$ either to obtain exactly 
$\Delta \chi^2 = t^2$ at $z_i=t$ for each of the $S_i^{\pm}$ sets or to get 
the best average agreement over some $z_i$ range (for instance for 
$z_i \le t$). Below we apply the second approach.


\subsubsection{Estimate of the tolerance parameter $T$ for the photon 
densities}

We consider now the value of the tolerance parameter $T$ for the real-photon 
parton-densities corresponding to the allowed deviation of the global fit from
the minimum, $\Delta \chi^2 = T^2$, as described above. In case of an ideal 
analysis a standard requirement is $\Delta \chi^2=1$. Of course this is not a 
case for a global fit to the $F_2^{ \gamma}$ data coming from various 
experiments, and certainly $T$ must be greater than 1. Unfortunately, no 
strict rules allowing for estimation of the tolerance parameter exist, as 
discussed in detail in \cite{cteq2} and \cite{cteq4}. We try to estimate the 
reliable $T$ value in two ways, both applied to the CJK 2 fit only.

First we examine the mutual compatibility of the experiments used in the fit.
That gives the allowed $\Delta \chi^2$ greater than 22 and the tolerance 
parameter $T \sim 5$. As a second test we compare the results of our four fits
presented in this paper and find the $T$ values for which parton densities 
predicted by the FFNS and CJK 1 models lie between the lines of uncertainties 
of the CJK 2 model. These values are large due to the differences between the 
gluon densities given by the CJK and FFNS models. The $T\sim 5$ when only the 
CJK 1 and 2 gluon distributions are compared. For quark distributions 
$T \le 7$ when we consider all models. Finally we estimate that the tolerance 
parameter $T$ should lie in the range $5\sim 10$.


\subsubsection{Tests of quadratic approximation}

For each CJK model we obtained a set of the $\{(v_{i})_k \}$ and $\{s_k\}$ 
values, with $i$ and $k=1,\cdots,4$ (since $d=4$ in CJK models). We used the 
iteration procedure from \cite{cteq1}. Further we adjusted the scaling factors
$\{s_k\}$ to improve the average quadratic approximation over the $z_k \le 5$ 
range.

\begin{figure}[t]
\hskip -0.2cm \includegraphics[scale=1.0]{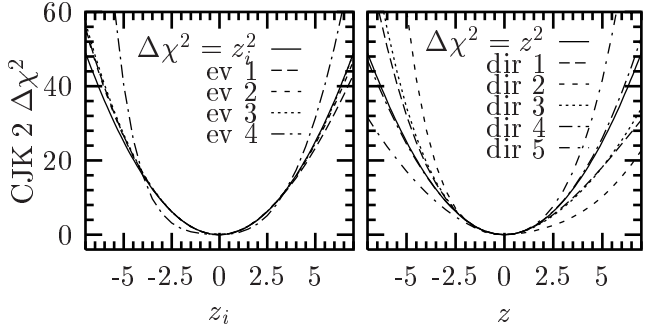}%
\vskip -0.8cm
\caption{CJK 2 model. The left plot presents a comparison of the $\chi^2$ 
dependence along each of the four eigenvector directions to the 
ideal $\Delta \chi^2 = z_i^2$ 
curve. In right plot analogous comparison for five random directions
in $\{z_i\}$ space are shown. The ideal curve corresponds to 
$\Delta \chi^2 = \sum z_i^2 = z^2$}
\label{quad11}
\end{figure}

Further we check if the quadratic approximation on which the Hessian method 
relies is valid in the considered $\Delta \chi^2$ range for the CJK 2 model
(for the CJK 1 model results are similar). In the left panel of Fig. 
\ref{quad11} we present the comparison of the $\chi^2$ dependence along each 
of four eigenvector directions (for the eigenvector $k$ $z_i=\delta_{ik}$) 
with the dependence of the ideal $\Delta \chi^2 = z_i^2$ curve. Only the line 
corresponding to the eigenvector 4 does not agree with the theoretical 
prediction. Moreover it has a different shape than other lines which results 
from the scaling adjustment procedure. In the right panel of Fig. \ref{quad11}
an analogous comparison for the five randomly chosen directions in the 
$\{z_i\}$ space is shown. For each of directions $\sum_{k=1}^dz_k^2=z^2$ and 
the ideal curve corresponds to $\Delta \chi^2 = z^2$. In this case we observe 
greater deviations from the quadratic approximation.

\begin{figure}[t]
\hskip -0.2cm \includegraphics[scale=0.7]{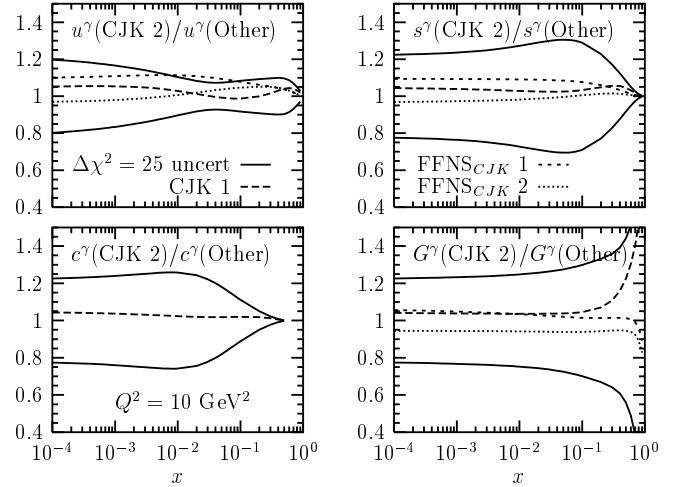}
\vskip -1.2cm
\caption{Parton densities calculated in FFNS$_{CJK}$ and CJK 1 models compared
with the CJK 2 predictions. We plot for $Q^2=10$ the 
$q\gam(\mathrm{CJK 2})/q\gam(\mathrm{Other})$ ratios of parton densities. 
Solid lines show the CJK 2 fit uncertainties for $\Delta \chi^2 = 25$.}
\label{densuncup}
\end{figure}


\subsubsection{CJK 1 and CJK 2 parametrizations}

The errors calculated within the Hessian quadratic approximation are shown in 
table \ref{parerrors} for the standard requirement of $\Delta \chi^2 = 1$. They
should be compared with the slightly smaller errors calculated by 
\textsc{Minuit} and shown in table \ref{tparam}.

\begin{table*}[htb]
\begin{center}
\renewcommand{\arraystretch}{1.2}
\begin{tabular}{|c|c|c|c|c|}
\hline
 model  & $\kappa$ & $\alpha$ & $\beta$ & $N_v$ \\
\hline
\hline
 CJK 1  & $2.146^{+0.120}_{-0.107}$  & $0.218^{+0.054}_{-0.047}$ & $0.462^{+0.157}_{-0.126}$ & $0.269^{+0.041}_{-0.035}$ \\
\hline
 CJK 2  & $1.934^{+0.112}_{-0.103}$  & $0.299^{+0.061}_{-0.051}$ & $0.898^{+0.204}_{-0.156}$ & $0.404^{+0.066}_{-0.054}$ \\
\hline
\end{tabular}
\caption{The parameters of the fits for CJK models with errors calculated in 
the Hessian quadratic approximation for the standard requirement of 
$\Delta \chi^2 = 1$}
\label{parerrors}
\end{center}
\end{table*}

All parton distributions are further parametrized on a grid. The resulting
programs can be found on the web-page \cite{webpage}.


\subsubsection{Uncertainties of the CJK 2 parton densities}

In this section we discuss the uncertainties of the CJK 2 parton 
densities, the results obtained with CJK 1 model are very similar.

In Figure \ref{densuncup} the up, strange and charm-quark and gluon densities 
calculated in FFNS$_{CJK}$ and CJK 1 models are compared with the CJK 2 
predictions. We plot the 
$q\gam(\mathrm{CJK \:2})/q\gam(\mathrm{Other \: models})$
ratios of the parton $q\gam$ densities calculated in the CJK 2 model to their 
values obtained with other models for $Q^2=10$. Solid lines show the CJK 2 fit
uncertainties for $\Delta \chi^2 = 25$ computed with the $\{S_k^{\pm}\}$ 
parametrizations.

First we notice that predictions of all our models for all considered parton 
distributions lie between the lines of the CJK 2 uncertainties. This indicates
that the choice of $\Delta \chi^2 = 25$ agrees with the differences among our 
four models. We found, that the SaS1D results differ very substantially from 
the CJK 2 ones (not shown).

As expected the up-quark distribution is the one best constrained by the
experimental data while the greatest uncertainties are for the gluon densities.
In the case of $u\gam$ the $\Delta \chi^2 = 25$ band widens in the small $x$ 
region. Alike in the case of $s$- and $c$-quark uncertainties it shrinks 
at high $x$. On contrary the gluon distributions are least constrained at the 
region of $x\to 1$. All uncertainties become slightly smaller for 
higher $Q^2$.


\subsection{Lagrange method for the uncertainties of the parton distributions}

The Hessian method allows the computation of the parton density uncertainties 
in a very simple and effective way. However, the Hessian method relies on the 
assumption of the quadratic approximation, which as we have shown in the 
former section, is not perfectly preserved.

There exist another method called the Lagrange multiplier method which allows 
to find exact uncertainties independently on the quadratic approximation (for
the proton structure used in \cite{cteq1},\cite{mrst} and \cite{cteq4}).
In this approach one makes a series of fits on the quantity
\be
F(\lambda,\{a_i\}) = \chi^2(\{a_i\}) + \lambda X(\{a_i\}),
\end{equation}
each with a different but fixed value of the Lagrange multiplier $\lambda$. As
a result one obtains a set of points $(\chi^2(\lambda),X(\lambda)$ which 
characterize the deviation of the physical quantity $X$ from its best value 
$X_0$, for a corresponding deviation of the structure function global fit from 
its minimum $\Delta \chi^2 = \chi^2(\lambda) - \chi_0$. In each of this 
constrained (by the $\lambda$ parameter) fits we find the best value of $X$ 
and the optimal $\chi^2$. For $\lambda = 0$ we return to the basic fit which 
gives $\{a_i^0\}$ parameters and allows to calculate the best $X_0$ value. The
great advantage of this approach lies in the fact that we do not assume 
anything about the uncertainties. The large computer time consuming of the 
process of the whole series of minimalizations is a huge disadvantage of the 
Lagrange method.


\subsection{Examples of cross-section uncertainties in Hessian and Lagrange 
methods}

Finally we made a comparison of the uncertainties obtained for the CJK 2 model 
in Hessian and Lagrange methods for two physical quantities. For the sake of 
limitation of the  computer time we chose two very simple examples: 
$F\gam_{2,c}$ points measured by the OPAL Collaboration \cite{F2c} and the 
$\gamma q \to \gamma q$ part of the cross-section for prompt photon production 
in $\gamma \gamma$. Results for the later case are presented in Fig. 5.


\section{Summary}

We enlarged and improved our previous analysis \cite{cjkl}. We performed new 
global fits to the \fun data. Two additional models were analysed. New fits 
gave $\chi^2$ per degree of freedom, 1.5-1.7, about 0.25 better than the old 
results. All features of the CJKL model, such as heavy-quark distributions, 
good description of the LEP data on the $Q^2$ dependence of the $F_2\gam$ and 
on $F_{2,c}\gam$ are preserved. We checked that the gluon densities of our 
models agree with the H1 measurement of the $G^{\gamma}$ distribution 
performed at $Q^2=74$ GeV$^2$ \cite{h1glu}, Fig. \ref{gluh1}.
\footnote{Further comparison of our gluon densities to the H1 data 
cannot be performed in a fully consistent way, since the GRV LO proton and 
photon parametrization were used in the experiment in order to extract such 
gluon density.}

An analysis of the uncertainties of the CJK parton distributions due to the 
experimental errors based on the Hessian method was performed for the very 
first time for the photon. We constructed sets of test parametrizations for 
both CJK models. They allow to computate uncertainties of any physical 
quantity depending on the real photon parton densities.

Parametrization programs for all models can be obtained from the web-page
\cite{webpage}.


\section{Acknowledgment}

This work was partly supported by the European Community's Human Potential 
Programme under contract HPRN-CT-2000-00149 Physics at Collider
and HPRN-CT-2002-00311 EURIDICE.


\begin{figure}[t]
\begin{center}
\input{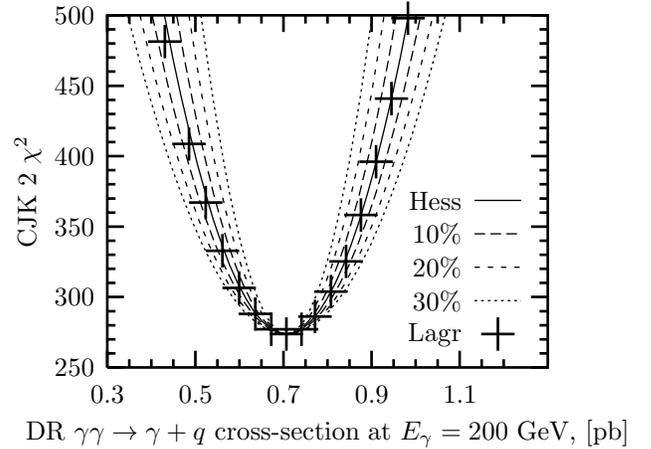} 
\vskip -0.5cm
\caption{CJK 2 model. Lagrange and Hessian method results for the direct 
resolved (DR) part of the $\gamma \gamma \to \gamma q$ cross-section. The 
dashed lines represent the 10 to 30\% deviation from the Hessian result.}
\end{center}
\label{fprompt}
\end{figure}

\begin{figure}[b]
\includegraphics[scale=1.0]{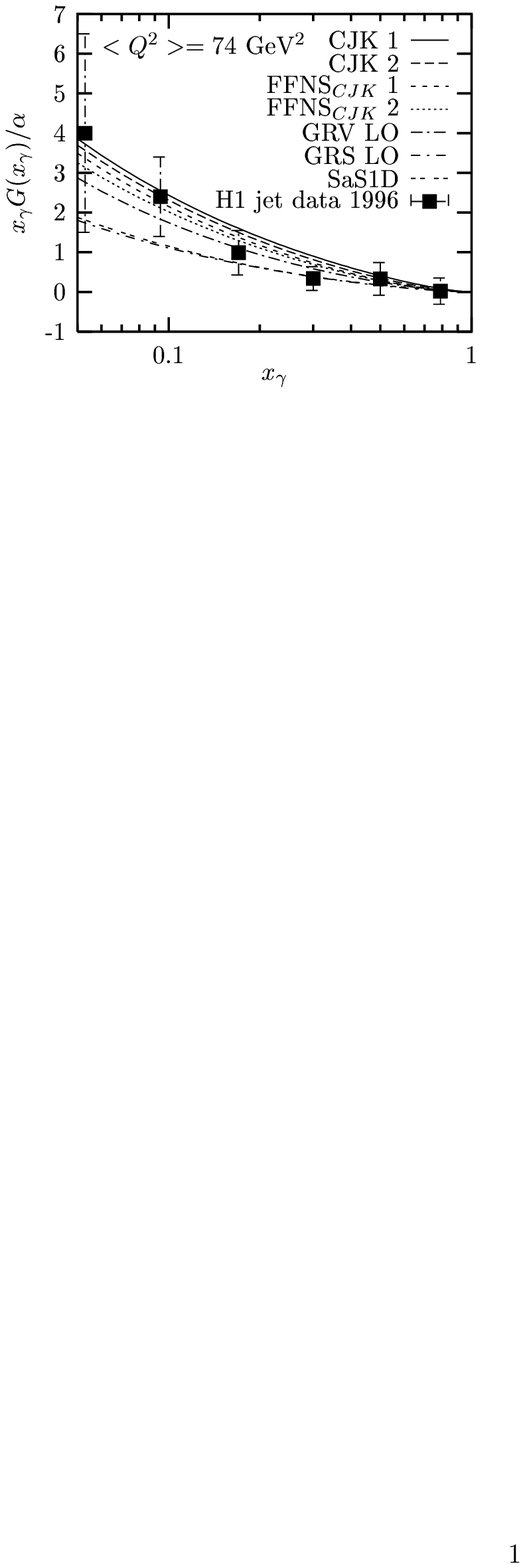}
\vskip -1.3cm
\caption{Comparison of the gluon distribution obtained in the H1 measurement 
performed at $Q^2=74$ GeV$^2$ \cite{h1glu} with the predictions of the CJK, 
FFNS$_{CJK}$ models and GRV LO \cite{grv92}, GRS LO \cite{grs} and SaS1D 
\cite{sas} parametrizations with the OPAL measurement \cite{F2c}.} 
\label{gluh1}
\end{figure}


\begin{thebibliography}{99}

\bibitem{cjkl} F. Cornet, P. Jankowski, M. Krawczyk and A. Lorca, 
               Phys. Rev. {\bf D68} 014010 (2003)
\bibitem{acot} S. Kretzer, C. Schmidt and W. Tung, 
               J. Phys. {\bf G28}, 983 (2002)
\bibitem{grv92} M. Gl\"uck, E. Reya and A. Vogt, 
                Phys. Rev. {\bf D46}, 1973 (1992)
\bibitem{cteq1} J. Pumplin, D.R. Stump and W.K. Tung, 
                Phys. Rev. {\bf D65}, 014011 (2002)
\bibitem{cteq2} J. Pumplin \etal, Phys. Rev. {\bf D65}, 014013 (2002)
\bibitem{cteq3} J. Pumplin \etal, JHEP 0207, 012 (2002)
\bibitem{mrst} A.D. Martin, R.G. Roberts, W.J. Stirling and R.S. Thorne, 
               Eur. Phys. J. {\bf C28}, 455 (2003)
\bibitem{cteq4} D. Stump \etal, Phys. Rev. {\bf D65}, 014012 (2002)
\bibitem{GRSt} M.Gl\"uck, E.Reya and M.Stratmann, 
               Phys. Rev. {\bf D51}, 3220 (1995)
\bibitem{klasen} S. Albino, M. Klasen and S. S\"oldner-Rembold, 
                 Phys. Rev. Lett. {\bf 89}, 122004 (2002).
\bibitem{prd} Particle Data Group (D.E. Groom \etal), 
              Eur. Phys. J. {\bf C15}, 1 (2000)
\bibitem{MariuszPrzybycien}We thank Mariusz Przybycie\'n for pointing this 
problem to us.
\bibitem{minuit} F. James and M. Roos, 
                 Comput. Phys. Commun. {\bf 10}, 343 (1975)
\bibitem{grs} M. Gl\"uck, E. Reya and I. Schienbein, 
              Phys. Rev. {\bf D60}, 054019 (1999), 
              Erratum-ibid. {\bf D62}, 019902 (2000)
\bibitem{sas} G.A. Schuler and T. Sj\"ostrand,
              Z. Phys. {\bf C68}, 607 (1995); 
              Phys. Lett. {\bf B376}, 193 (1996)
\bibitem{HQ2} OPAL Collaboration, G.~Abbiendi \etal, 
              Phys.\ Lett.\ {\bf B533}, 207 (2002).
\bibitem{F2c} OPAL Collaboration, G. Abbiendi \etal,
              Phys. Lett. {\bf B539}, 13 (2002)
\bibitem{newart} P. Jankowski PhD thesis, 
                 and F. Cornet, P. Jankowski, M. Krawczyk, in preparation 
\bibitem{webpage} http://www.fuw.edu.pl/\verb+~+pjank/param.html
\bibitem{h1glu} H1 Collaboration, C. Adloff \etal, 
                Phys. Lett. {\bf B483}, 36 (2000)
\end{thebibliography}
\end{document}